\documentclass[twocolumn,showpacs,amsmath,amssymb,prl]{revtex4}
\usepackage{graphicx}
\usepackage{dcolumn}
\usepackage{bm}

\begin{document}
\input{epsf}

\title{Possibility of Precise Measurement of the Cosmological Power\\
Spectrum With a Dedicated 21cm Survey After Reionization} \author{Abraham
Loeb$^{1}$ \& J. Stuart B. Wyithe$^2$} \affiliation{$^1$ Astronomy
Department, Harvard University, 60 Garden Street, Cambridge, MA 02138, USA}
\affiliation{$^2$School of Physics, University of Melbourne, Parkville,
Victoria, Australia}

\begin{abstract}

Measurements of the 21cm line emission by residual cosmic hydrogen after
reionization can be used to trace the power spectrum of density
perturbations through a significant fraction of the observable volume of
the Universe. We show that a dedicated 21cm observatory coule probe a
number of independent modes that is two orders of magnitude larger than
currently available, and enable a cosmic-variance limited detection of the
signature of a neutrino mass $\sim 0.05$ eV.  The evolution of the linear
growth factor with redshift could also constrain exotic theories of gravity
or dark energy to an unprecedented precision.

\end{abstract}


\date{\today}

\maketitle

Recently, there has been much interest in the feasibility of mapping the
three-dimensional (3D) distribution of cosmic hydrogen through its resonant
spin-flip transition at a rest-frame wavelength of
21cm\cite{Furlanetto,BarLoeb}. Several experiments are currently being
constructed (such as
MWA\footnote{http://www.haystack.mit.edu/ast/arrays/mwa/},
LOFAR\footnote{http://www.lofar.org/}, PAPER
\footnote{http://astro.berkeley.edu/~dbacker/EoR/},
21CMA\footnote{http://web.phys.cmu.edu/~past/}) and more ambitious
designs are being planned (SKA\footnote{http://www.skatelescope.org/})
to detect the theoretically-predicted emission signal.

Measurements of the power-spectrum of 21cm brightness fluctuations could
constrain the initial conditions from inflation as well as the nature of
the dark matter and dark energy.  The 21cm fluctuations are expected to
simply trace the primordial power-spectrum of matter density perturbations
either before the first galaxies had formed (at redshifts $z\gtrsim
20$)\cite{LZ04,Lewis} or after reionization ($1\lesssim z\lesssim 6$) --
when only dense pockets of self-shielded hydrogen (such as damped
Ly$\alpha$ systems) survive \cite{WL07,Pen}.  During the epoch of
reionization, the fluctuations are mainly shaped by the topology of ionized
regions \cite{McQuinn,Trac,Iliev}, and thus depend on astrophysical
details.  However, even during this epoch, the line-of-sight anisotropy of
the 21cm power spectrum due to peculiar velocities, can in principle be
used to separate the implications for fundamental physics from the unknown
details of the astrophysics \cite{BL04,McQuinn}.  In what follows, we will
focus our discussion on the post-reionization epoch \cite{WL07,Pen} which
offers two advantages. First, it is least contaminated by the Galactic
synchrotron foreground (whose brightness temperature scales with the
redshift under consideration as $(1+z)^{2.6}$ \cite{Furlanetto}). Second,
because the UV radiation field is nearly uniform after reionization, it
should not imprint any large-scale features on the 21cm power spectrum that
would mimic cosmological signatures.  On large spatial scales the 21cm
sources are expected to have a linear bias analogous to that inferred from
galaxy redshift surveys.  Since a 21cm survey maps the global hydrogen
distribution without resolving individual galaxies, the 21cm bias is
expected to be modest compared to surveys that select for the brightest
galaxies at the same redshifts.

In general, cosmological surveys are able to measure the power-spectrum of
primordial density fluctuations, $P(k)$, to a precision that is ultimately
limited by cosmic variance, namely the number of independent Fourier modes
that fit within the survey volume. 21cm observations are advantageous
relative to existing data sets because they access a 3D volume instead of
the 2D surface probed by the cosmic microwave background (CMB), and they
extend to a sufficiently high redshift (well beyond the horizon of galaxy
redshift surveys \cite{SDSS}) where most of the comoving volume of the
observable Universe resides.  At these high redshifts, small-scale modes
are still in the perturbative (linear growth) regime where their analysis
is straightforward.  The expected 21cm power extends down to the
pressure-dominated (Jeans) scale of the cosmic gas which is orders of
magnitude smaller than the comoving scale at which the CMB anisotropies are
damped by photon diffusion \cite{LZ04}.

Altogether, the above factors make 21cm surveys an ideal cosmological probe
of fundamental physics \cite{Kl}.  To illustrate this point, we show in
this {\it Letter} that a dedicated 21cm observatory would enable a
determination of the matter power-spectrum at redshifts $z\lesssim 6$ to an
unprecedented precision. In our numerical examples, we adopt the standard
set of cosmological parameters \cite{Spergel}.

\begin{figure} [ht]
\centerline{\epsfxsize=3.2in \epsfbox{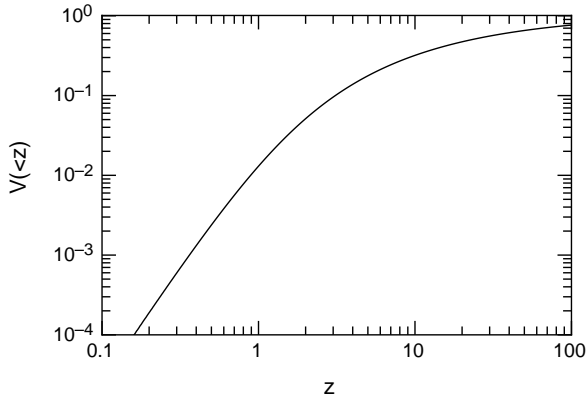}} 
\caption{The fraction of the total comoving volume of the observable
Universe that is available up to a redshift $z$.}
\label{fig1}
\end{figure}

\paragraph*{Number of Modes.}
The limitation of existing redshift surveys of galaxies
\cite{Tegmark,Lahav} is apparent in Fig.~\ref{fig1} which plots the
comoving volume of the Universe out to a redshift $z$ as a function of $z$.
State-of-the-art galaxy redshift surveys, such as the spectroscopic sample
of luminous red galaxy (LRGs) in the {\it Sloan Digital Sky Survey (SDSS)}
\cite{SDSS}, extend only out to $z\sim 0.5$ (over $\sim10\%$ of the sky)
and probe $\sim 0.01\%$ of the observable Universe.

The CMB fluctuations probe a thin shell on the 2D surface of the sky.  The
number of modes with a comoving wave number $k\equiv2\pi/\lambda$ between
$k$ and $k+dk$ that fit on this 2D surface is, $dN_{\rm CMB} = \pi k dk
\left[\mathcal{A}/(2\pi)^2\right]$, where $\mathcal{A}=D^2d\Omega$, $D$ is
the comoving distance to the last scattering surface at $z\sim 10^3$ and
$d\Omega$ is the solid angle of the survey field. Redshift surveys of
galaxies and 21cm surveys probe a 3D comoving volume $\mathcal{V}$ and
potentially access a larger number of modes, $dN_{\rm 3D} = 2\pi k^2 dk
\left[\mathcal{V}/(2\pi)^3\right]$.

\begin{figure} [ht]
\centerline{\epsfxsize=3.2in \epsfbox{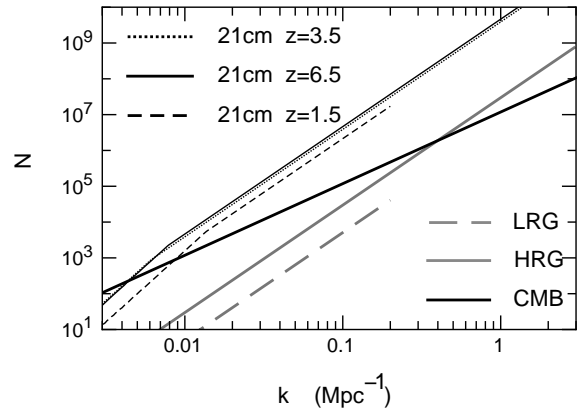}} 
\caption{The number of modes $N$ within a wave number bin of width $\Delta
k=k/10$ centered on $k$, that are available in different cosmological
surveys. The thick dashed grey line corresponds to the spectroscopic LRG
sample of {\it SDSS} \cite{SDSS}, while the thick solid line (marked HRG)
corresponds to a future spectroscopic survey at $2.5<z<3.5$ covering 1000
square degrees with a a co-moving galaxy density equal to the LRG
sample. The thick dark line corresponds to a CMB data set with $f_{\rm
sky}=0.65$. The thin lines show the number of modes accessible in a 21cm
survey (including the limit on large scale modes due to foreground removal
\cite{McQuinn}) covering $f_{\rm sky}=0.65$ within a redshift range
spanning a factor of 3 in $(1+z)$, and centered on $z=1.5,3.5$ and
$6.5$. For $z\leq1.5$, we have truncated the curves at $k=0.2~{\rm
Mpc^{-1}}$ to illustrate the smaller range of $k$ accessible within the
linear regime at lower redshifts \cite{eisen}. }
\label{fig2}
\end{figure}

Figure~\ref{fig2} compares $N_{\rm CMB}=(k/10)\,dN_{\rm CMB}/dk$ with
$N_{\rm 3D}=(k/10)\,dN_{\rm 3D}/dk$, for future 21cm surveys after
reionization \cite{WL07}.  The CMB data set is assumed to cover a fraction
$f_{\rm sky}=0.65$ of the sky (excluding the region around Milky Way
galaxy). For comparison, we also show the corresponding number of modes
within the same $k$ interval in the spectroscopic LRG sample of {\it SDSS}
\cite{SDSS}, which covers $\sim3700$ square degrees out to $z\sim0.5$ or a
volume of $\mathcal{V}=0.72h^{-3}$Gpc$^3$ (where $h\approx0.7$ is the
Hubble constant in units of 100${\rm km~s^{-1}~Mpc^{-1}}$).

21cm observatories that are currently under construction (such as MWA) will
survey only a few percent of the sky and process only $\sim15$\% of the
available frequency range (band-pass). In this {\it Letter}, we consider
future 21cm surveys that would potentially cover $f_{\rm sky}=0.65$ with a
processed frequency-bandwidth spanning a redshift range of a factor of 3 in
$(1+z)$ centered on $z=1.5$, $z=3.5$ and $z=6.5$ \footnote{The factor of 3
in $(1+z)$ corresponds to the largest frequency bandwidth over which a
low-frequency dipole antenna has suitable sensitivity.}.  With an array
design based on MWA in which the effective area of each tile of 16
dipole antennae equals its physical area, the value of $f_{\rm sky}=0.65$
corresponds to $\sim33$ correlated primary beams or fields.

\begin{figure} [ht] 
\centerline{\epsfxsize=3.2in \epsfbox{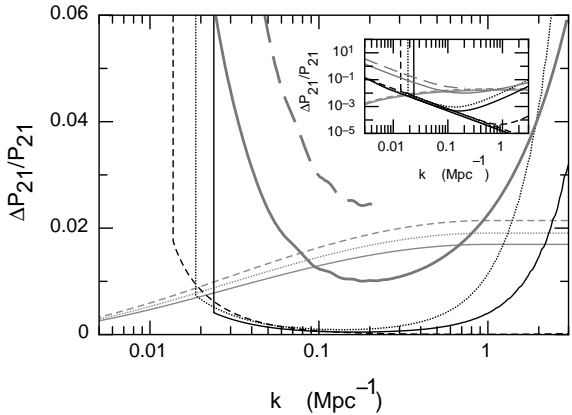}}
\caption{ The fractional change in the amplitude of the power-spectrum
owing to the presence of a massive neutrino (horizontal grey lines,
asymptoting towards a constant at high $k$ values). The case shown,
$f_\nu=0.004$, corresponds to $m_\nu=0.05$eV.  For comparison, the limits
imposed by cosmic variance on measurements of the power-spectrum from the
{\it SDSS}-LRG and a future 1000 square degree galaxy survey at $2.5<z<3.5$
are marked by the thick dashed and solid grey lines respectively. The {\it
U}-shaped error curves correspond to an all-sky 21cm survey [$f_{\rm
sky}=0.65$ over a redshift range spanning a factor of 3 in $(1+z)$] with
MWA5000 and a $10^3$ hour integration per field (line styles for
$z=1.5,3.5,6.5$ as in Figure~\ref{fig2}). The noise is plotted in bins of
size $\Delta k/k = 0.1$. The inset shows these results on logarithmic axes
that span a larger dynamic range of achievable precision.  The straight
thin lines in the inset show the cosmic-variance uncertainty in the
power-spectrum measurement owing only to the number of available modes. 
For $z\leq1.5$, we have truncated the curves at $k=0.2~{\rm
Mpc^{-1}}$ to illustrate the smaller range of $k$ accessible within the
linear regime at lower redshifts \cite{eisen}. }
\label{fig3}
\end{figure}

\paragraph*{Results.}  The fractional uncertainties in the 21cm
power-spectrum $P_{21}$ for a cosmic-variance limited survey, $(\Delta
P_{21}/P_{21})=1/\sqrt{N}$, are presented in the inset of Fig.~\ref{fig3}
(straight lines). Also shown in both the inset and main panel are the noise
curves for observations using a design based on the so-called MWA5000
experiment, which is assumed to have 10 times the collecting area of MWA
(as described in Refs. \cite{McQuinn,WLG}). MWA5000 would be cosmic
variance limited in an integration time of $\sim 10^3$ hours at wave
numbers near $k\sim0.1$Mpc$^{-1}$.  Since the dipoles of each antenna tile
look at $\sim \pi$ steradians, simultaneous processing of multiple primary
beams would allow a survey of area $f_{\rm sky}=0.65$ at $10^3$ hours of
integration per pointing within a few years.  In computing the thermal
noise of the observatory, we have adopted a model for the bias factor
$b_{21}$ of the 21cm sources ($b_{21}^2\equiv P_{21}/P$) from
Ref. \cite{WL07}.  The noise curves include a limit on large scale modes
due to foreground removal. Current estimates for MWA show that foreground
removal should be effective for modes over a frequency range $\lesssim
6$MHz which is ${1\over 4}$ of the total 24MHz processed bandwidth
\cite{McQuinn}. However, improvements on this range would provide access to
a larger $N$ as well as to lower-$k$ modes. In Fig.~\ref{fig3} and
subsequently, we assume a scenario in which the foreground can be removed
on scales of up to ${1\over 12}$ of the total processed bandwidth, namely
${1\over 12}\times (\sqrt{3}-1/\sqrt{3})[1400{\rm MHz}/(1+z)]=
30[(1+z)/4.5]^{-1}{\rm MHz}$.
 
For comparison, we also show the noise curves for the {\it SDSS}-LRG survey
(thick grey line), including the effects of Poisson shot-noise\footnote{We
ignore Poisson fluctuations in the 21cm power spectra since the
contributing galaxies are expected to be of much lower mass than LRGs, and
there should be a large number of them per resolution element of the
survey.}, $(\Delta P_{\rm gal}/P_{\rm gal})=[1+(b^2P(k)n_{\rm
gal})^{-1}]/\sqrt{N}$, for a galaxy number density of $n_{\rm
gal}=46748/(0.72h^{-3})$Gpc$^{-3}$ and a bias factor of $b_{\rm gal}=2$
\cite{SDSS}.  We find that the potential 21cm constraints on the matter
power-spectrum are 1--2 orders of magnitude better than a low-redshift
galaxy survey like {\it SDSS}-LRG.

As an example for the potential use of a 21cm survey, we show in
Fig.~\ref{fig3} the expected relative changes in the amplitude of the
power-spectrum owing to the presence of a massive neutrino \cite{EH}.  At
wave numbers much larger than the neutrino free-streaming wave number
($k_{\rm fs}= 0.1\Omega_m(0)h\sqrt{f_{\nu}}~{\rm Mpc^{-1}}$), the
suppression of the power-spectrum is given by \cite{neutrino,Hannestad},
\begin{equation}
\frac{P(k,f_\nu)}{P(k,f_{\nu}=0)}=(1-f_\nu)^3\left[1.9\times10^5
g(0)\Omega_m(0){f_\nu \over N_\nu}\right]^{-6f_\nu/5},
\end{equation} 
where $g(z)\approx \Omega_m^{0.2}/[1+0.003(\Omega_\Lambda/\Omega_m)^{4/3}]$
is the growth function of the gravitational potential for matter and
vacuum density parameters of
$\Omega_m(z)=[1+\Omega_\Lambda/\Omega_m(0)(1+z)^3]^{-1}$ and
$\Omega_\Lambda=(1-\Omega_m)$ \cite{KGB}; $f_\nu\equiv
[\Omega_\nu(0)/\Omega_m(0)] =0.08 N_\nu (m_\nu/1~{\rm eV})$ is the
present-day mass fraction of the matter density carried by $N_\nu$ neutrino
species of particle mass $m_\nu$. In this {\it Letter} we conservatively
assume a non-degenerate hierarchy of neutrino masses with $N_\nu=1$ and
$m_\nu$ denoting the largest mass eigenstate.  Figure~\ref{fig3} shows the
case corresponding to $f_\nu=0.004$
($m_\nu=0.05$eV).  

While Fig.~\ref{fig3} indicates that the cosmic variance in a galaxy
redshift survey (such as the {\it SDSS}-LRG survey) is sufficiently small
to detect the suppression of power at $k\gg k_{\rm fs}$ due to a neutrino
mass of $m_\nu=0.05$eV, uncertainties in other cosmological parameters and
parameter degeneracies reduce this sensitivity by an order of
magnitude\cite{EH,Slosar}.  Thus, in order to avoid possible systematic
offsets between the power-spectrum amplitude observed by different
techniques (such as galaxy surveys, CMB maps, and Ly$\alpha$ forest data)
at different $k$ values, it is desirable for the 21cm survey to be
self-contained and cover a sufficiently broad range of wave numbers that
probe the curvature of the neutrino effect in Fig. \ref{fig3}.  For this to
be achieved, the removal of the Galactic synchrotron foreground would need
to be effective over large frequency intervals of up to $\sim
30[(1+z)/4.5]^{-1}{\rm MHz}$.  Foreground removal across such intervals
would provide access to the required range of scales over which a constant
offset (in the form of a linear bias factor) would not affect the $m_\nu$
measurement. The desired wave number to be reached by foreground removal
corresponds to the scale where cosmic variance is larger than the change
induced by a massive neutrino.  Currently, the detailed properties of the
foreground are not well measured. The feasibility of foreground removal
over a broad frequency interval will remain uncertain until data from the
first generation of 21cm observatories is analysed.

Figure~\ref{fig3} illustrates that foreground removal to $\sim
30[(1+z)/4.5]^{-1}{\rm MHz}$ would be sufficient to detect the modification
of the power spectrum due to the minimum neutrino mass of $0.047\pm 0.01$eV
indicated by the latest atmospheric neutrino data \cite{neutrino,Maltoni}
at all spatial scales where the effect is larger than cosmic variance. The
advantage of the large survey volume is evident since it allows the
modification of shape to be measured in addition to the suppression
detected by the {\it SDSS}-LRG survey (the latter being cosmic variance
limited on scales where the shape is measured).  In a follow-up paper
\cite{Visb}, we will address the level of degeneracy with other
cosmological parameters.  Already, Fisher-matrix studies \cite{Mao,McQuinn}
have demonstrated the improved capabilies of 21cm observations during the
epoch of reionization, where contamination from astrophysical sources needs
to be removed through the angular dependence of the 21cm power-spectrum
\cite{BL04}.

A 21cm survey measures the modulation in the cumulative 21cm emission from a
large number of galaxies, as its coarse angular and redshift resolution
is not capable of resolving the 21cm sources individually \cite{WL07,WLG}.
The damped Ly$\alpha$ systems which contain most of the hydrogen mass in
the Universe at $z\lesssim 6$, are expected to be hosted by abundant low
mass galaxies \cite{Wolfe} and thus have a weak bias relative to the
underlying matter distribution on large spatial scales. This weak bias is
not expected to introduce a feature to the power-spectrum that is
degenerate with the neutrino signature (as would be the case prior to
reionization).  Although the comoving wave number at which non-linear
evolution becomes important increases from $k\sim0.1h^{-1}$Mpc$^{-1}$ at
$z=0.3$ to $\sim0.5h^{-1}$Mpc$^{-1}$ at $z=3$ \cite{eisen}, the constraints
on $m_\nu$ can be potentially improved by accounting for the related
non-linear effects \cite{Saito}.


By measuring the evolution of the growth factor with redshift to the
exquisite precision implied by Fig. \ref{fig3}, it would also be possible
to constrain alternative theories of gravity or dark energy well beyond the
capabilities of existing data sets \cite{Hui}. 
The evolution of the growth factor would be limited by the knowledge of the
bias of the 21cm sources, $b_{21}$. The limit on the uncertainty in growth
factor would satisfy $[d\ln{g}/d\ln{(1+z)}]\approx
[d\ln{b_{21}}/d\ln{(1+z)}]$.

Finally, we note that a precise $P(k)$ measurement at multiple redshifts
would also allow to determine the redshift evolution of the baryonic
acoustic oscillations (BAO) in the 21cm power spectrum \cite{BL05,WL07}.
The BAO scale constitutes a standard ruler \cite{eisen,BlGl} that can be
used to measure the equation of state of the dark energy \cite{WLG,Pen},
constrain $1-\Omega_{\rm tot}$, and further remove degeneracies between
$m_\nu$ and other cosmological parameters \cite{Wong}.

\paragraph*{Hardware.} 
The required 21cm observatory could be similar in antenna design to the
planned MWA but would require expanded collecting area and computer
resources to account for the increased cross-correlation requirements and
the analysis of multiple beams. A suitable observatory would contain ten
times the number of tiles in MWA.  The computational load increases as the
square of the number of tiles in a telescope, and linearly with both the
amount of processed band-pass and the number of cross-correlated primary
beams. Thus, an all-sky survey over a frequency range covering a factor of
3 in $(1+z)$ requires an overall improvement by $\sim 10^4$ in computer
power relative to MWA\footnote{The cost of the antenna hardware is expected
to be in the range of hundreds of million of dollars, an order of magnitude
higher than for MWA. Improved telescope designs \cite{Mao} and advances in
computer technology could help to realize our projected 21cm observatory.}.

\bigskip
\paragraph*{Acknowledgments.}
We thank M. McQuinn and M. Zaldarriaga for helpful comments.  We also thank
the Harvard-Australia foundation for funding a visit of A.L.  to Australia,
during which this work was performed.

\end{document}